\documentclass[aps,prl,twocolumn,superscriptaddress]{revtex4-1}

\usepackage{graphicx}
\usepackage{natbib}
\usepackage[breaklinks]{hyperref}
\usepackage{amsmath}

\newcommand{\ket}[1]{|{#1}\rangle}

\newcommand{\hflev}[4]{\textit{#1}$_{#2/#3}$, \textit{F}=#4}

\begin{document}

\title{Time-resolved Scattering of a Single Photon by a Single Atom}
\author{Victor Leong}
\affiliation{Center for Quantum Technologies, 3 Science Drive 2, Singapore 117543}
\affiliation{Department of Physics, National University of Singapore, 2 Science Drive 3, Singapore 117542}
\author{Mathias Alexander Seidler}
\affiliation{Center for Quantum Technologies, 3 Science Drive 2, Singapore 117543}
\author{Matthias Steiner}
\affiliation{Center for Quantum Technologies, 3 Science Drive 2, Singapore 117543}
\affiliation{Department of Physics, National University of Singapore, 2 Science Drive 3, Singapore 117542}
\author{Alessandro Cer\`{e}}
\affiliation{Center for Quantum Technologies, 3 Science Drive 2, Singapore  117543}
\author{Christian Kurtsiefer}
\affiliation{Center for Quantum Technologies, 3 Science Drive 2, Singapore 117543}
\affiliation{Department of Physics, National University of Singapore, 2 Science Drive 3, Singapore 117542}
\date{\today}



\maketitle
\textbf{Scattering of light by matter has been studied extensively in the past. 
  Yet, the most fundamental process, the scattering of a single  photon by a
  single atom, is largely
  unexplored~\cite{Piro:2010js,Sandoghdar:2012,Brito:2016}. One prominent
  prediction of quantum optics is the deterministic absorption of a traveling
  photon by a single atom, provided the photon waveform matches spatially and 
  temporally the time-reversed version of a spontaneously emitted
  photon~\cite{Quabis:2000,Imamoglu:2008,Leuchs:2009,Wang:2011,Bader:2013,
    Heugel:2009,Syed:2013,Liu:2014,Srivathsan:2014jx}. Here, we experimentally
  address this prediction and investigate the influence of the temporal
  profile of the photon on the scattering dynamics using a single trapped atom
  and heralded single photons. In a time-resolved measurement of the atomic
  excitation we find a~56(11)\% increase of the peak excitation  by photons
  with an exponentially rising profile compared to a decaying one. This result
  demonstrates that tailoring the envelope of single photons enables precise
  control of the photon-atom interaction.
  }

The efficient excitation of atoms by light is a prerequisite for many proposed quantum information protocols. 
Strong light-matter interaction by using either large ensembles of
atoms~\cite{Hammerer2010,Lukin2003} or single atoms inside
cavities~\cite{Gleyzes:2007jr,Wilk:2007hm,Stute:2012de} has received much
attention in the past. More recently, significant light-matter interaction has also
been observed between single quantum systems and weak coherent fields in free
space~\cite{Tey:2008,Slodicka:2010,Streed:2012,Fischer:2014}. 
The time-reversal symmetry of Schroedinger's and Maxwell's equations suggests that the conditions for perfect absorption of an incident single
photon by a single atom in free space can be found from the reversed process,
the spontaneous emission of a photon from an atom prepared in an excited state. 
There, the excited state population decays exponentially with a time constant given by the radiative lifetime~$\tau_0$ of the excited state, and an outward-moving photon with the same temporal
decay profile emerges in a spatial field mode corresponding to the atomic dipole transition~\cite{Weisskopf:1930jm}. 
Therefore, for efficient atomic excitation the incident photon should have an
exponentially rising temporal envelope with a matching time constant~$\tau_0$ and propagate in the atomic dipole mode towards the position of the atom~\cite{Leuchs:2012ps}. 

For a more quantitative description of the scattering process we follow  Ref.~\cite{Wang:2011}, which assumes a stationary two-level atom interacting with a propagating single photon in the Weisskopf-Wigner approximation. 
The photon-atom interaction strength depends on the spatial
overlap~$\Lambda\in [0,1]$ of the atomic dipole mode with the propagating mode of the photon, where $\Lambda = 1$ corresponds to complete spatial mode overlap.
In this work, we consider scattering of exponentially decaying and rising photons described by the probability amplitude~$\xi(t)$ 
 \begin{equation}\label{eq:photons}
     \xi_{ \downarrow}(t) = \frac{1}{\sqrt{\tau_p}}\Theta(t)
     e^{-\frac{t}{2\tau_p}}
\quad{\rm and}\quad
     \xi_{ \uparrow}(t) = \frac{1}{\sqrt{\tau_p}}\Theta(-t) e^{\frac{t}{2\tau_p}}\,,
 \end{equation}
where $\Theta(t)$ is the Heaviside step function and
${\tau_p}$ is the coherence time of the photon.
Integrating the equations of motion in Ref.~\cite{Wang:2011} leads to analytic
expressions for the time-dependent population~$P_{e}(t)$ in the excited state of the atom
for both photon shapes:
\begin{equation}\label{eq:P_e_down}
  P_{e,{ \downarrow}}(t)=
    \begin{cases}
        \frac{4\Lambda\tau_0 \tau_p}{\left(\tau_0-\tau_p\right)^2}  \Theta(t)\left( e^{ -\frac{t}{2\tau_0}  } -e^{ -\frac{t}{2\tau_p }  }\right)^2& \text{ for }\tau_p \neq \tau_0 \\
        \frac{ \Lambda t^2}{{\tau_0}^2} \Theta(t) e^{ -\frac{ t}{\tau_0} } & \text{ for } \tau_p =\tau_0
      \end{cases}
\end{equation}
and
\begin{equation}\label{eq:P_e_up}
     P_{e,\uparrow}(t)= \frac{4\Lambda\tau_0 \tau_p}{\left(\tau_p+\tau_0\right)^2} \left [ e^{ \frac{ t}{\tau_p} }\, \Theta(-t) +   e^{- \frac{ t}{\tau_0} }\, \Theta(t) \right]\,.
\end{equation}

\begin{figure}[ht]
\centering
\includegraphics[width=\columnwidth]{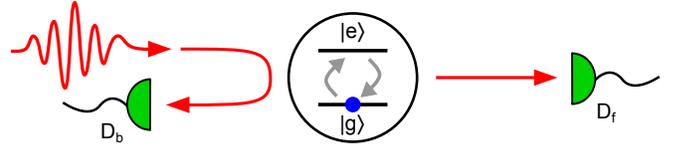}
  \caption{\label{fig:idea}
  An incident single photon excites a two-level atom in free-space.
  The time evolution of the atomic excited state population can be inferred by measuring 
  photons in the forward or backward direction. 
   }
\end{figure}
In our experiment (Fig.~\ref{fig:idea}) we focus single probe
photons onto a single atom, and infer the atomic excited state population~$P_e(t)$ from photons arriving at the forward and backward detectors~$D_f$ and~$D_b$.
We obtain~$P_e(t)$ directly from the atomic fluorescence measured at the backward detector~D$_\textrm{b}$ with the detection probability per unit time~$R_{b} (t)$,
\begin{equation} \label{eq:back}
  P_e(t) =   \frac{\tau_0 } {\eta_b}R_{b}(t) \,
\end{equation}
where $\eta_b$ is the collection efficiency. 
However,
the detection rate in such an experiment is relatively small.
Alternatively, $P_e(t)$ can be determined from the detection rate at the
forward detector~D$_\textrm{f}$ with a better signal-to-noise ratio.
The probability per unit time of detecting a photon in the forward direction at time
$t$ is given by $R_{f,0}\left( t\right)=|\xi(t)|^2$ without an atom, and by
$R_{f}(t) = \left|\xi(t) - \sqrt{ \frac{\Lambda} {\tau_0} P_e(t)}\right|^2$ with an
atom present. 
The atom alters the rate of transmitted photons via absorption and re-emission towards the forward detector~D$_\textrm{f}$. 
Therefore, any change~$ \delta\left( t\right)$ of the forward detection rate is directly related to a change of the atomic population,
\begin{equation}\label{eq:delta}
  \delta\left( t\right) = R_{f,0}\left(t\right) - R_{f}\left( t\right).
\end{equation}
The excited state population~$P_e(t)$ is then obtained by integrating a rate equation,
\begin{equation} \label{eq:P_e_dot}
  \dot{P}_e(t) =  \delta(t) - \frac{(1 -\Lambda )}{\tau_0}  P_e(t) \,,
\end{equation}
where the last term describes spontaneous emission into modes that do not overlap with the excitation mode. 

\begin{figure}[t]
  \includegraphics[width=\columnwidth]{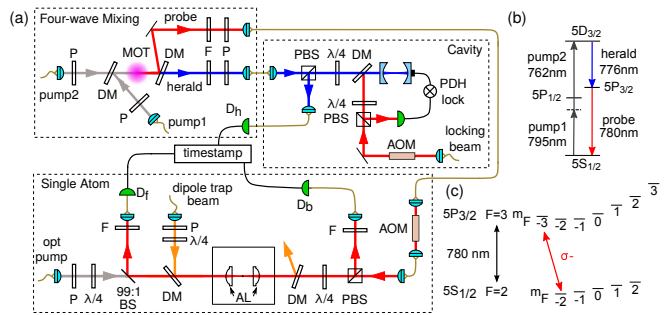}
  \centering
  \caption{\label{fig:joint_setup}
    (a) Experimental setup. (Top left) Four-wave mixing part, providing
    heralded single photons: Pump\,1 (795\,nm) and Pump\,2 (762\,nm)  are overlapped in a copropagating geometry inside the cold cloud of  $^{87}${Rb} atoms
    in a magneto-optical trap, generating pairs of herald (776\,nm) and probe (780\,nm) photons.
    The detection of a photon at D$_\textrm{h}$ heralds a probe photon.
    (Top right) Tuning the resonance of a bandwidth-matched cavity  with respect to the heralding photon frequency controls the temporal envelope.
    (Bottom) Single atom part: A $^{87}${Rb} atom is trapped at the focus of a confocal aspheric lens pair
    (AL; numerical aperture 0.55) with a far-off-resonant optical dipole trap ($980\,\text{nm}$). 
    The probe photons are guided to the single atom part by a single mode fiber and focused onto the atom by the first AL. 
    Avalanche photodetectors D$_\textrm{f}$ and D$_\textrm{b}$ detect photons collected in forward and backward directions.
    An acousto-optic modulator (AOM) shifts the probe photon frequency to compensate for the shift of the atomic resonance frequency caused by the bias magnetic field and the dipole trap. 
    D$_\textrm{h}$, D$_\textrm{f}$, D$_\textrm{b}$: avalanche photodetectors (APDs), P: polarizer, F: interference filters, $\lambda$/2, $\lambda$/4: half- and quarter-wave plates, (P)BS: (polarizing) beam splitter, DM: dichroic mirror.
    (b)~Relevant level scheme of the four-wave mixing process in a cloud of $^{87}${Rb} atoms. 
    (c)~Relevant level scheme of the single  $^{87}${Rb} atom in the dipole trap. The probe photons are resonant with the closed transition \mbox{$\ket{g}$\,=\,5\hflev{S}{1}{2}{2}, $m_F$=-2} to  \mbox{$\ket{e}$\,=\,5\hflev{P}{3}{2}{3}, $m_F$=-3}.
    }
\end{figure}
A schematic of the experimental setup is shown in Fig.~\ref{fig:joint_setup}.
A single $^{87}$Rb atom is trapped at the joint focus of an aspheric lens pair
(numerical aperture 0.55) with a far-off-resonant optical dipole trap
($980\,\text{nm}$)~\cite{Tey:2008}. After molasses cooling, the trapped atom
is optically pumped into the 5\hflev{S}{1}{2}{2}, $m_F$=-2 state.
Probe photons are prepared by heralding on one photon of a time-correlated photon pair generated via four-wave-mixing (FWM) in a cloud of cold $^{87}$Rb atoms~\cite{Chaneliere:2006,Srivathsan:2013}.
The relevant energy levels are depicted in Fig.~\ref{fig:joint_setup}(b):
two pump beams with wavelengths 795\,nm and 762\,nm excite the atoms from 5\hflev{S}{1}{2}{2} to 5\hflev{D}{3}{2}{3}, and
a subsequent ensemble-enhanced cascade decay gives rise to the time ordering necessary for obtaining exponential time envelopes~\cite{Franson:1992cl,Srivathsan:2014jx,Gulati:2014}.
Dichroic mirrors, interference filters and coupling into single mode fibers select photon pairs of wavelengths 776\,nm (herald) and 780\,nm (probe).
Adjusting the atomic density of the atomic ensemble~\cite{Srivathsan:2013}, we set the coherence time~$\tau_p=13.3(1)$\,ns of the generated photons,
corresponding to a spectral
overlap with the atomic linewidth of approximately~$90\%$~\cite{Leong:2015hom}.

To control the temporal envelope of the probe photon, the heralding mode is
coupled to a bandwidth-matched, asymmetric Fabry-Perot cavity. 
The cavity reflects the herald photons with a dispersive phase shift depending on the cavity resonance frequency. 
Tuning the cavity on resonance or far-off resonance (70\,MHz) with respect to
the center frequency of the herald photon results in exponentially rising or
decaying probe photons~\cite{Srivathsan:2014jx}.
The FWM source alternates between a laser cooling interval of~140\,$\mu$s, and
a photon pair generation interval of~10\,$\mu$s, during which we register on
average~0.054 heralding events on avalanche photodetector~(APD)~D$_\textrm{h}$.
The probe photons are guided to the single atom  by a single mode fiber.  
The spatial excitation mode is then defined by the collimation lens at the
output of the fiber and the  high numerical aperture aspheric lens AL. 
From the experimental geometry we expect a spatial mode overlap of
$\Lambda\approx0.03$ with the atomic dipole mode~\cite{Syed:2013}.
The excitation mode is then collimated by a second aspheric lens, again
coupled into a single-mode fiber, and sent to the forward
detector~D$_\textrm{f}$.
A fraction of the photons scattered by the atom is collected in the backward
direction, and similarly fiber-coupled and guided to detector~D$_\textrm{b}$.
\begin{figure}[th]
\centering
  \includegraphics[width=\columnwidth]{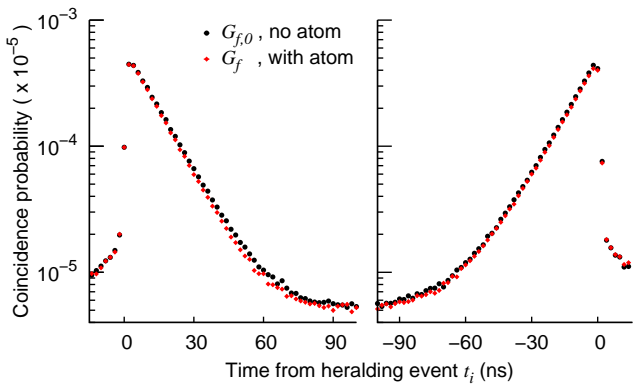}
  \caption{\label{fig:tx_w_wo_atom}
  Coincidence histograms between heralding detector~D$_\textrm{h}$ and forward
  detector~D$_\textrm{f}$ for exponentially decaying (left) and rising (right)
  probe photons with a coherence time~$\tau_p=13.3(1)$\,ns obtained from a fit.
  Black circles: $G_{f,0}$, reference data taken without the trapped atom. Red
  diamonds: $G_f$, data taken with the atom present.
  The time bin size is $\Delta t=2$\,ns. Total measurement time is~1500\,hours. Error bars are smaller than the symbol size. 
  We offset all detection times by 879\,ns to account for delays introduced by electrical and optical lines.
    }
\end{figure}

\begin{figure}[ht]
\centering
  \includegraphics[width=\columnwidth]{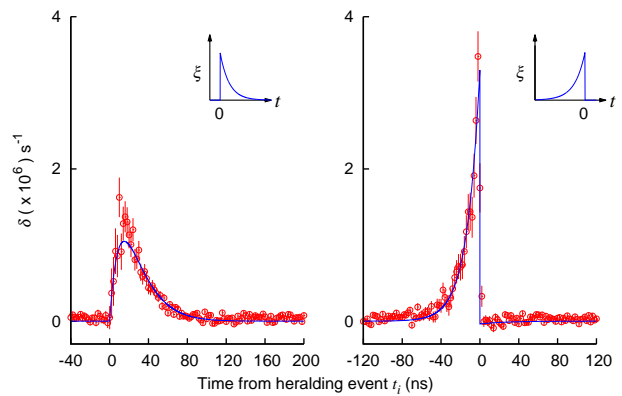}
  \caption{\label{fig:tx_diff}
  Changes in the forward detection rates~$\delta\left( t_{i}\right) =
  R_{f,0}\left( t_{i}\right) -R_{f}\left( t_{i}\right)$ induced by the interaction with the atom. 
  The time bin size is 2\,ns. 
  Solid lines: analytical solution using
  Eq.~(\ref{eq:P_e_down}-\ref{eq:delta}) for $\tau_p=13.3\,$ns, $\Lambda=0.033$. 
  Left and right columns show results for exponentially decaying and rising probe photons, respectively.
    }
\end{figure}

To investigate the dynamics of the scattering process, we record photoevent
detection times at the forward detector~D$_\textrm{f}$ with respect to
heralding events at~D$_\textrm{h}$.
When no atom is trapped, we obtain the reference
histograms~$G_{f,0}\left(t_i\right)$ for exponentially decaying and rising
probe photons, with time bins $t_i$ of width~$\Delta
t$~(Fig.~\ref{fig:tx_w_wo_atom}, black circles).
The observed histograms resemble closely the ideal asymmetric exponential
envelopes, described by Eq. \ref{eq:photons}.
The total probability of a coincidence event within a time interval of
114\,ns ($\approx$ 8\,$\tau_p$) is $\eta_f=3.70(1)$\,$\cdot10^{-3}$.
When an atom is trapped, we record histograms~$G_{f}(t_i)$~(Fig.~\ref{fig:tx_w_wo_atom}, red diamonds). 
The two histograms~$G_{f}\left(t_i\right)$ are very similar to the respective reference histograms~$G_{f,0}(t_i)$. 
To reveal the scattering dynamics we obtain the photon detection probabilities per unit time at the forward detector~$R_f(t_i) = G_f(t_i)/(\eta_f \Delta t)$
with and without atom in order to use Eq.~(\ref{eq:delta}-\ref{eq:P_e_dot}) to reconstruct the excited state population~$P_e(t_i)$.  
Figure~\ref{fig:tx_diff} shows the difference~$\delta(t_i)=
R_{f,0}(t_i)-R_f(t_i)$ for both photon envelopes, with mostly positive values.
A positive value of $\delta\left( t_{i}\right)$ corresponds to net absorption,
i.e., a reduction of the number of detected photons during the time bin $t_i$
due to the interaction with the atom.
For a photon with a decaying envelope, the absorption is close to zero at $t_i=0$, and reaches a maximum at $t_i\approx 15$\,ns, followed by a slow decay.
In strong contrast, the absorption for photons with a rising envelope follows
the exponential envelope of the photon, with a maximum absorption rate twice
as high as that for photons with a decaying envelope.
We find that the magnitude and the dynamics of the observed scattering is well reproduced by Eq.~(\ref{eq:P_e_down}-\ref{eq:delta}) for $\tau_p=13.3\,$ns and $\Lambda=0.033$~(Fig.~\ref{fig:tx_diff},  solid lines).

The interaction with the atom reduces the overall transmission into the forward detection path for both photon shapes.
To quantify this behavior, we calculate the extinction~$\epsilon=\Delta
t\sum_i \delta(t_i)$ by summing over the interval $-14\,\textrm{ns}\leq
t_{i} \leq 100\,\textrm{ns}$ for exponentially decaying photons, and
$-100\,\textrm{ns} \leq t_i \leq 14\,\textrm{ns}$ for exponentially rising
photons, capturing almost the entire photon.
We obtain similar extinction values $\epsilon_\downarrow = 4.21\,(18) \%$ and $\epsilon_\uparrow = 4.40\,(20) \%$ for decaying and rising photons, respectively.
The theoretical value of the extinction does not depend on whether the photon envelope is exponentially decaying or rising: 
\begin{equation}
\epsilon =  \int^{+\infty}_{-\infty} \delta(t) dt= \Lambda
  \left( 1- \Lambda \right) \frac{4\tau_p}{\tau_0+\tau_p}
\end{equation}
For our parameters, $\tau_p=13.3\,$ns, $\Lambda=0.033$,
this expression leads to $\epsilon=4.29\%$, which is close to our experimental results. 
\begin{figure}[ht]
\centering
  \includegraphics[width=\columnwidth]{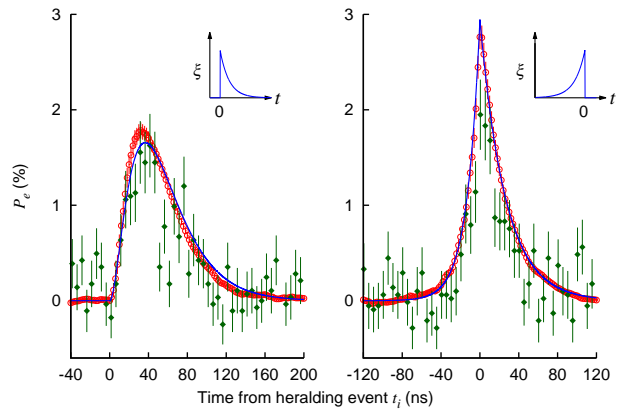}
  \caption{\label{fig:p_e}
  Atomic excited state population obtained from the forward (red open circles,
  time bin size 2\,ns) and backward detection rates (green filled diamonds,
  time bin size 5\,ns). Solid lines: $ P_e(t) $ from Eq.~(\ref{eq:P_e_down}) and~(\ref{eq:P_e_up})
    using $\tau_p=13.3\,$ns, $\Lambda=0.033$.
  Left and right columns show results for exponentially decaying and rising probe photons, respectively.
    }
\end{figure}

The excitation probability~$P_e(t_i)$ (Fig.~\ref{fig:p_e}, red circles) of the
atom is obtained from the differences in the forward detection
rates~$\delta(t_i)$ and by numerically integrating Eq.~(\ref{eq:P_e_dot}). 
The exponentially decaying photon induces a longer lasting but lower atomic
excitation compared to the rising photon. 
We find good agreement with the analytical solutions given in Eq.~(\ref{eq:P_e_down}) and~(\ref{eq:P_e_up}) (Fig.~\ref{fig:p_e},  solid line).
We do not observe perfect excitation of the atom from exponentially rising probe photons 
because of the small spatial mode overlap~$\Lambda$. 
However, the peak excited state population for the exponentially rising~$P_{e,\max,\uparrow}=2.77(12)\%$ is $56(11)\%$ larger than for the decaying one~$P_{e,\max,\downarrow}=1.78(9)\%$. 
The increase in the peak excitation~\mbox{$P_{e,\uparrow,\max}/P_{e,\downarrow,\max}=78\%$} predicted by  Eq.~(\ref{eq:P_e_down}) and~(\ref{eq:P_e_up}) for $\tau_p=13.3\,$ns, $\Lambda=0.033$ is also in fair agreement with our findings.

The excited state population can also be directly determined from the atomic fluorescence, Eq.~(\ref{eq:back}).
To convert the coincidence histograms~$G_b(t_i)$ between the heralding detector~D$_\textrm{h}$ and backward detector~D$_\textrm{b}$ 
into the excited state population~$P_e(t_i)$ we have to account for the finite collection and detection efficiencies in the forward and backward path. 
For the backward path we independently  measure the collection efficiency~$\eta_b=0.0126(5)$ and the detector quantum efficiency~$\eta_q=0.56(1)$.  
Figure~\ref{fig:p_e} (green filled diamonds) shows the inferred excited state population~$P_e(t_i) = R_{b}(t_i)/ (\eta_b \Gamma_0) = G_b(t_i)/( \tilde{\eta}_f \eta_q \eta_b  \Gamma_0 \Delta t) $ 
with a time bin width of 5\,ns,
where $\tilde{\eta}_f=0.0155(4)$ is the heralding efficiency in the forward
path, corrected for the collection and detection efficiencies. 
Again, we find a qualitatively different transient atomic excitation for both
photon shapes, in agreement with the theoretical model, but with worse detection
statistics compared to the excited state reconstruction using the changes in the forward detection rates.

In summary, we have accurately measured the atomic excited state population during photon scattering and have demonstrated that the power spectrum of the incident photon is not enough to fully characterize the interaction. 
The exponentially rising and decaying photons have an identical Lorentzian power spectrum with a full-width-half-maximum $\Gamma_p=\frac{1}{\tau_p}$, but the transient atomic excitation differs. 
We have shown that the scattering dynamics depends on the envelope of the photon, in particular that an atom is indeed more efficiently excited by a photon with an exponentially rising temporal envelope compared to an exponentially decaying one. 
However, when integrated over a long time interval $\Delta t \gg \tau_0,\tau_p$ both photon shapes are equally likely to be scattered as shown by our measurement of the extinction~$\epsilon$.
The advantage of using exponentially rising photons is, therefore, to excite atoms at well defined instants in time.
Such a synchronization can be beneficial to quantum networks. 

Our experimental results also contribute to a longstanding
discussion about differences between heralded and ``true'' single photons. The
atomic excitation dynamics caused by heralded single photons matches well the  
one expected from ``true'' single photon states in our theoretical model,
and therefore support a realistic interpretation of photons prepared in a
heralding process. 

We acknowledge the support of this work by the Ministry of Education in
Singapore (AcRF Tier 1) and the National Research Foundation, Prime
Minister's office (partly under grant No. NRF-CRP12-2013-03). 
M. Steiner acknowledges support by the Lee Kuan Yew Postdoctoral Fellowship.

\textbf{Methods}

\textbf{Heralded single photon generation:}
The two pump fields have orthogonal linear polarizations.
The 795\,nm pump laser is red-detuned 
by~$-30$\,MHz from the 5\hflev{S}{1}{2}{2} to 5\hflev{P}{1}{2}{2} transition to avoid incoherent scattering. 
The frequency of the 762\,nm pump laser is set such that the two-photon
transition from 5\hflev{S}{1}{2}{2} to 5\hflev{D}{3}{2}{3} is driven with a
blue-detuning of 4\,MHz.
We can vary the coherence time~$\tau_p$ of the generated photons by changing the optical density of the atomic ensemble. 
We choose~$\tau_p=13.3$\,ns as a trade-off between matching the excited state
lifetime of $\tau_0=26.2$\,ns and having a high photon pair generation rate.
Longer coherence times can be achieved at lower optical densities, but at the cost of lower photon pair generation rates.  
 
The probe photons are guided to the single atom setup by a 230\,m long optical fiber. 
An acousto-optic modulator~(AOM) compensates for the 72\,MHz shift of the atomic resonance frequency caused by the bias magnetic field (7~Gauss applied along the optical axis) and the dipole trap. 
The AOM also serves as an optical switch between the two parts of the experimental setup; 
once a herald photon is detected, the AOM is turned on for~$600$\,ns. 
The optical and electrical delays are set such that the probe photon passes the AOM within this time interval. 
Before reaching the atom, the polarization of the probe photons is set to
circular~$\sigma^-$ by a polarizing beam splitter and a half-wave-plate.

The Fabry-P\'{e}rot cavity used to control the temporal envelope has a length of 125\,mm and a finesse of $103(5)$, resulting in a decay time $\tau_c=13.6 (5)$\,ns. 
The reflectance of the in-coupling mirror and the second mirror are~$0.943$ and~$0.9995$ respectively.  
We use an auxiliary 780\,nm laser to stabilize the cavity length using the Pound-Drever-Hall technique. 

\textbf{Data acquisition and analysis:}
Fig.~\ref{fig:tx_w_wo_atom} shows the coincidence histograms without additional processing, 
while corrections for accidental coincidences were applied in the analyzed data shown in Fig.~\ref{fig:tx_diff} and Fig.~\ref{fig:p_e},

The total acquisition time for the experiment was~1500\,hours, during which the average photon coherence time was~$\tau_p=13.3(1)$\,ns and the heralding efficiency was~$\eta_f=3.70(1)$\,$\cdot10^{-3}$. 
We check for slow drifts in  $\tau_p$ and $\eta_f$ by analyzing the histogram~$G_{f,0}$ every 60\,min for $\tau_p$ and 20\,min for $\eta_f$.
The distribution of $\tau_p$ is nearly Gaussian with a standard deviation of
$0.9$\,ns, most likely caused by slow drifts of the laser powers and the
atomic density; the distribution of $\eta_f$ is slightly asymmetric with a full-width-half-maximum of~$6$\,$\cdot10^{-4}$.
We alternated between the decaying and rising photon profiles every 20~min to
ensure that the recorded coincidence histograms are not systematically biased
by slow drifts in $\tau_p$ and $\eta_f$.

\bibliographystyle{apsrev4-1}
%

\end{document}